\documentclass[12pt]{iopart}
\usepackage{bbm}
\usepackage{iopams}
\eqnobysec

\newcommand{\refeq}[1]{(\ref{#1})}%%
\newcommand{\tfrac}[2]{\frac{#1}{#2}}
\newcommand{\rmT}  {{\mathrm{T}}}
\newcommand{\rmVM}  {{\mathrm{VM}}}
\newcommand{\rmHM}  {{\mathrm{HM}}}

\newcommand{\Vol}  {\mathrm{Vol}}

\newcommand{\const}  {\mathrm{const}}

\newcommand{\eg} {{\it e.g.}}

\newcommand{\Exp}[1]{\exp\left\{{\big.{#1}}\right\}}
\newcommand{\yt}{{\tilde{y}}}
\newcommand{\xiT}{\tilde{\xi}}

\newcommand{\rmdH}{\hat{\rmd}}
\newcommand{\Rset}{\mathbbm{R}}

\newcommand{\Ac} {\mathcal{A}}

\newcommand{\Fc} {\mathcal{F}}

\newcommand{\Mc}{\mathcal{M}}
\newcommand{\Nc}{\mathcal{N}}

\newcommand{\Uc}{\mathcal{U}}

\newcommand{\Pc}{\mathcal{P}}

\newcommand{\Vc}{\mathcal{V}}
\newcommand{\OmegaB}{\bar{\Omega}}
\newcommand{\Fcb}{\bar{\Fc}}
\newcommand{\Fb}{\bar{F}}

\newcommand{\Wb}{\bar{W}}
\newcommand{\Yb}{\bar{Y}}
\newcommand{\Zb}{\bar{Z}}
\newcommand{\zb}{\bar{z}}

\newcommand{\jb}{{\bar{\jmath}}}
\newcommand{\hb}{{\bar{h}}}

\newcommand{\nablab}{\overline{\nabla}}
\newcommand{\PdS}{\big(\vec{\Pc}\cdot\vec{\sigma}\big)}
\newcommand{\nS}{\big(\vec{n}\cdot\vec{\sigma}\big)}
\newcommand{\PP}{\big(\vec{\Pc}\cdot\vec{\Pc}\big)}
\newcommand{\eL}{e_I L^I}
\newcommand{\eY}{\big(e_I Y^I\big)}

\def\der{\partial}

\def\fffrac#1/#2{\leavevmode\kern.1em
  \raise.5ex\hbox{\the\scriptfont0 {#1}}\kern-.1em
  /\kern-.15em\lower.25ex\hbox{\the\scriptfont0 {#2}}}

\def\ffrac#1/#2{\leavevmode\kern.1em
  \raise.5ex\hbox{$#1$}\kern-.1em
  \Big/\kern-.15em\lower.25ex\hbox{$#2$}}

\newcommand{\ft}[2]{{\textstyle\frac{#1}{#2}}}
\begin{document}
%%\renewcommand{\baselinestretch}{1.5}\normalsize
%%     \hfill hep-th/0407198\\
%%     \hfill FSU-TPI-05/04\\
\title{Domain Walls, Hitchin's Flow Equations and $G_2$-Manifolds%%%
%%%  \footnote{Work supported by the `Schwerpunktprogramm Stringtheorie' of the DFG.}
  }

\author{Christoph Mayer and  Thomas Mohaupt}
\address{Theoretisch-Physikalisches Institut,
  Friedrich-Schiller-Universit\"{a}t
  Jena, Max-Wien-Platz 1, D-07743 Jena, Germany
}
 
\eads{\mailto{Christoph.Mayer@uni-jena.de}, \mailto{Thomas.Mohaupt@uni-jena.de}}

\begin{abstract}
  We construct BPS domain wall solutions of the effective
  action of type-IIA string theory compactified on a half-flat 
  six-manifold. The flow equations for the vector and hypermultiplet
  scalars are shown to be equivalent to Hitchin's flow equations,
  implying that our domain walls can be lifted to solutions of ten-dimensional
  type-IIA supergravity. They take the  form $\mathbbm{R}^{1,2} \times Y_7$, 
  where  $Y_7$ is a $G_2$-holonomy manifold with boundaries.
\end{abstract}

\submitto{\CQG}
\maketitle

%%%%%%%%%%%%%%%%%%%%%%%%%%%%%%%%%%%%%%%%%%%%%%%%%%%%%%%%%%%%%%%%%%%%%%%%%%%%%%%%
\section{Introduction}
%%%%%%%%%%%%%%%%%%%%%%%%%%%%%%%%%%%%%%%%%%%%%%%%%%%%%%%%%%%%%%%%%%%%%%%%%%%%%%%%

Type-II string compactifications on Calabi-Yau three-folds have been
studied extensively in the past, because they provide deep insights
into the dynamics of four-dimensional string theories. We refer
to  \cite{Greene} for a review and references. One particularly fascinating and
useful property
is mirror symmetry \cite{MSBook}, which relates type-IIA string theory on a
Calabi-Yau manifold $X_6$ to type-IIB string theory on a different
Calabi-Yau manifold $\tilde{X}_6$, called the mirror manifold.
However, type-II Calabi-Yau compactifications do not lead
to realistic phenomenology. In particular, all massless modes
are gauge singlets, and one has a moduli space of degenerate vacua. 
This situation improves when
one considers more general type-II backgrounds, where vacuum 
expectations values of tensor fields are turned on 
\cite{PolchinskiStrominger, TaylorVafa, D'A2, D'A3, BC, CKLS}.
The back-reaction of the background flux deforms the geometry, 
which no longer is Ricci-flat and Calabi-Yau. The most popular approach 
to these compactifications
is based on the effective supergravity description and assumes
that the flux and its back-reaction can be treated as a 
perturbation away from a given Calabi-Yau compactification. This way 
one obtains a consistent and physically plausible picture:
switching on fluxes in the internal space corresponds 
to specific gaugings of the four-dimensional
effective supergravity theory. Some of the fields become
charged, and a scalar potential is created, which lifts some of
the flat directions. Moreover, this treatment also makes
sense geometrically: if one imposes that the internal space $X_6$ still
supports eight Killing spinors, it must be a manifold with 
$SU(3)$ structure \cite{Joyce, ChiossiSalomon, CardosoFiveT, GMW, GLMW}. 
Like 
Calabi-Yau manifolds,
which have $SU(3)$ holonomy,
manifolds with $SU(3)$ structure are characterised
by the existence of a non-degenerate real two-form $J$ and of a
complex three-from $\Omega$. However, these forms need not be 
closed, but satisfy weaker requirements \refeq{su3eq}.
Therefore, such manifolds are the natural generalisation 
of Calabi-Yau manifolds. 
Recently, Hitchin \cite{HitchinGenComplex} has introduced the notion of a `generalised
complex manifold', which generalises complex and symplectic geometry. 
For applications to flux compactifications and to mirror symmetry see
\cite{LMTZ,OBB,Jeschek,GMPT}. 

One natural question is whether mirror symmetry still holds in this
enlarged class of compactifications. For compactifications with
RR fluxes a mirror map between the corresponding 
four-dimensional effective supergravity actions has been found in 
\cite{LM}.
Mirror symmetry acts in a natural way on the RR fluxes by exchanging
the roles of RR tensor fields of even and odd rank. When considering
mirror symmetry for compactifications with NS fluxes, one 
immediately encounters
the question of what is the mirror partner of the NS flux.
In the context of the topological string it was observed that
switching on NS flux in type-IIB corresponds to deforming 
the type-IIA Calabi-Yau three-fold into a manifold with a 
non-integrable complex structure \cite{Vafa}.
A detailed proposal was made by \cite{GLMW}
who argued that the mirror partner of type-IIB string theory
on a (conformally rescaled) 
Calabi-Yau three-fold with NS flux
is type-IIA string theory on a so-called half-flat six-manifold. 
In this particular case, the back-reaction of the flux on the 
geometry only amounts to a conformal rescaling of the Calabi-Yau metric \cite{GPP,GP,D'A3}.
Conversely, type-IIA string theory on a Calabi-Yau three-fold with NS flux was shown to be mirror
symmetric to type-IIB string theory on a half-flat six-manifold \cite{GM}.
Half-flat manifolds are manifolds with a particular type of $SU(3)$ structure,
see \refeq{half-flat-Eq}. 
Unfortunately, the mathematical deformation theory  of 
manifolds with $SU(3)$ structure has not yet been developed to the same
degree as the one of Calabi-Yau manifolds, see, however, \cite{GIP}. 
Nevertheless, \cite{GLMW}
succeeded in employing
supersymmetry and mirror symmetry to motivate a particular
parametrisation of the differential forms occurring in
the dimensional reduction on half-flat six-manifolds 
(see Eqs.~\refeq{hf_basis_i},~\refeq{hf_basis_d}).
Using this, they obtained a mirror symmetric four-dimensional
supergravity action.
%
%with a gauging which is determined by
%the parameters which control the deformation of a Calabi-Yau
%three-fold into a half-flat six-manifold.
The gauging is determined by the parameters which control the deformation of a Calabi-Yau
three-fold into a half-flat six-manifold.

There is, however, one aspect of the proposal of \cite{GLMW}
which deserves further investigation. On the type-IIA side no
fluxes are turned on, and half-flat six-manifolds are not Ricci-flat.
Therefore $\mathbbm{R}^{1,3} \times X_6$ is not a solution to the
ten-dimensional type-IIA equations of motion.
By mirror symmetry,
the same is true on the type-IIB side.
What does it mean then
to perform a dimensional reduction of the action on $X_6$?
In principle, the answer has already been given in \cite{GLMW}.
The four-dimensional gauged supergravity action does not have
maximally supersymmetric Minkowski or AdS vacua. One expects, however,
from general experience with gauged supergravity actions
that it has BPS domain wall solutions with four Killing spinors.
In such solutions some of the scalar fields  have a non-trivial
dependence on the transverse coordinate of the domain wall.
If all these scalars can be interpreted as moduli of an internal 
space, the four-dimensional solution should lift
to a ten-dimensional one where the internal six-manifold varies
along the transverse direction. The correct question is whether
this geometry satisfies the ten-dimensional equations of motion.

The purpose of this paper is to check this explicitly. 
In Section 2 we construct new domain wall solutions of gauged
four-dimensional supergravity, by extending the results of Ref.~\cite{BCL} to
the case of an arbitrary hypermultiplet sector. In particular, 
we find that the vector multiplet sector can be treated without
specifying the gauging in the hypermultiplet sector. When analysing
the hypermultiplet sector we consider the generalised axion gauging \refeq{killingVs}
occurring in 
the dimensional reduction of type-IIA string theory on half-flat
six-manifolds. 
This gauging is different from the axion gauging considered in
\cite{BCL}.
In Section 3 we relate these four-dimensional domain walls
to ten-dimensional geometries. The domain walls are 
completely determined by flow equations, which 
specify how the scalars evolve as functions of the transverse
coordinate of the domain wall. Moreover, all scalars with 
a non-trivial flow come from internal components of the
ten-dimensional type-IIA metric, {\it i.e.} they are moduli of
the internal manifold. We then show that the scalar flow equations
are equivalent to Hitchin's flow equations. This implies that 
the internal half-flat six-manifold varies along the transverse direction
of the domain wall precisely in such a way that they combine
into a seven-manifold $Y_7$ with $G_2$ holonomy. Moreover,
the four-dimensional domain wall lifts to the ten-dimensional
space-time $\mathbbm{R}^{1,2} \times Y_7$ which is manifestly Ricci-flat and
therefore satisfies the type-IIA equations of motion. This clarifies
the meaning of the four-dimensional action obtained in \cite{GLMW}:
it describes the dynamics of the lightest modes around this
background. 
In Ref.~\cite{BDKT} a similar geometry has been investigated, where 
in a non-compact setup explicit expressions for the metrics on $Y_7$ and and $X_6$ have
been obtained. 
Section 4 contains an outlook onto further directions of research.

%%%%%%%%%%%%%%%%%%%%%%%%%%%%%%%%%%%%%%%%%%%%%%%%%%%%%%%%%%%%%%%%%%%%%%%%%%%%%%%%
\section{Domain-Wall Solutions}
%%%%%%%%%%%%%%%%%%%%%%%%%%%%%%%%%%%%%%%%%%%%%%%%%%%%%%%%%%%%%%%%%%%%%%%%%%%%%%%%
In this section, we construct domain-wall solutions of four-dimensional gauged
$\Nc$~$=$~$2$ supergravity \cite{deWvP}. 
Our analysis extends the work of Ref.~\cite{BCL} to the case
of arbitrary hypermultiplet gaugings.
Then we specialise to a particular
hypermultiplet gauging, which arises in compactifications of type-IIA 
supergravity on
so-called half-flat manifolds \cite{GLMW}.

The supergravity theory contains a gravity multiplet  
$\big\{g_{\mu\nu},\; \psi_{\Lambda\mu},\; \Ac^0_\mu\big\}$,
$n_V$~vector multiplets $\big\{\Ac_\mu^i,\; \lambda^{i\Lambda},\; z^i\big\}_{i=1}^{n_V}$, 
and
$n_H$~hypermultiplets
$\big\{q^u,\; \xi_\alpha\big\}_{u=1}^{4n_H}$.

We follow the conventions of
Refs.~\cite{BCL,sugra4d}. Space-time indices are denoted $\mu=0,1,2,3$, and
the signature of 
$g_{\mu\nu}$ is $\{+1,-1,-1,-1\}$.
All spinors are Majorana. The index $\Lambda=1,2$ is an $SU(2)$ index, and the index
$\alpha$ transforms in the fundamental representation of $Sp(2n_H)$. The
position of the indices $\Lambda$, $\alpha$ on spinors encodes 
a chiral projection \cite{sugra4d}.

The scalars $\{z^i,q^u\}$ are coordinates on $\Mc_\rmVM\times\Mc_\rmHM$, 
which are
special K\"ahler \cite{deWvP2} and quaternionic-K\"ahler
manifolds \cite{BagWit}, respectively. Special K\"ahler 
manifolds are characterised by the existence of a holomorphic
prepotential for the metric, which is 
homogeneous of degree two. Quaternionic-K\"ahler manifolds 
of real dimension $4n_H$ 
have a holonomy group which is contained in $SU(2) \times
Sp(2n_H)$. The indices $\Lambda,\alpha$ are tangent space
indices with respect to $\Mc_\rmHM$. Note that $Sp(2n_H)$ denotes
the compact real form of the symplectic group.
Since loop corrections to higher-dimensional 
quaternionic-K\"ahler manifolds are not yet accessible,\footnote{
See however \cite{AMT} for the case of the universal hypermultiplet.}
we work at string tree level and take $\Mc_\rmHM$ 
to be in the image of the c-map \cite{c-map,FS}.
Then $\Mc_\rmHM$ is determined by a special K\"ahler 
manifold
and thus can be specified by a prepotential.
This special K\"ahler manifold is the vector
multiplet manifold of a T-dual type-II compactification, 
and not to be confused with $\Mc_\rmVM$.

For the vector multiplet geometry we take
the prepotential to be of the form
\begin{equation}\label{Fc}
  \Fc(Y)= \frac{c_{ijk}Y^iY^jY^k}{Y^0}\;\quad z^i:= Y^i/Y^0\;,
\end{equation}
where $z^i$ are special coordinates \cite{deWvP2}.
The fields $Y^I$ are part of the symplectic vector 
$(Y^I ,\, \Fc_I)^\rmT$, $I=0\dots n_V$,
where $\Fc_I = \der_I \Fc(Y)$. Geometrically, the prepotential
$\Fc(Y)$ defines a `conic holomorphic non-degenerate
lagrangian immersion' of $\Mc_\rmVM$ into the projectivisation
$P\big(T^*\mathbbm{C}^{n_V +1}\big)$ of the symplectic 
vector space 
$T^*\mathbbm{C}^{n_V +1}$, and the components
of $(Y^I ,\, \Fc_I)^\rmT$ are the canonical embedding coordinates \cite{ACD}. 
In the context of
Calabi-Yau compactifications it is useful to consider
$(Y^I ,\, \Fc_I)^\rmT$ as the holomorphic section of a symplectic 
vector bundle over $\Mc_\rmVM$ \cite{Str}.
Note that not all such sections come from a prepotential. 
However, one can always find a symplectic transformation, such that a
prepotential exists \cite{vP}. 
%This vector is a section of a 
%a symplectic vector bundle over $\Mc_\rmVM$ which can be used to define 
%special
%K\"ahler manifolds. Sometimes one uses ``flat'' sections 
%$V$~$=$~$\begin{pmatrix}L^I\\M_I\end{array}\right)$~$:=$~$\Exp{-\Kc/2}\Omega$, 
%where $\Kc$ is the K\"ahler potential of $\Mc_\rmVM$.
%Locally there are so-called special coordinates in which 
%$\Fc_I=\partial_I\Fc(Y)$ and
%$Y^I=Y^I(z)$.

In type-IIA Calabi-Yau compactifications, a cubic prepotential
(\ref{Fc}) corresponds to working in the large-radius limit, where quantum 
corrections to $\Fc(Y)$ due to world-sheet
instantons are small and can be ignored.
The numbers $c_{ijk}$ are the triple intersection
numbers of the Calabi-Yau threefold.

We work at string tree-level, so that the quaterion-K\"ahler manifold $\Mc_\rmHM$ is
obtained by applying the c-map to a 
special K\"ahler manifold %$\Mc_\mathrm{SK}$
\cite{FS}. 
The resulting parametrisation of $\Mc_\rmHM$ is
%%%%%%%%%%%%%%%%%%%%%%%%%%%%%%%%%%%%%%%%%%%%%%%%%%%%%%%%%%%%%%%%%%%%%%%%%%%%%%%%
%% old version
%%%%%%%%%%%%%%%%%%%%%%%%%%%%%%%%%%%%%%%%%%%%%%%%%%%%%%%%%%%%%%%%%%%%%%%%%%%%%%%%
%%\begin{equation}\label{hyperVariables}
%% \big\{q^u\big\}_{u=1}^{4n+4} := \Big(V,a,\xi^A,\xiT_A, z^a,\zb^a\Big)\;,\quad
%%  A\in\big\{0\dots n\big\}\;,\quad n:=n_H-1\,,
%%\end{equation}
%%where $z^a$, $a\in\{1\dots n\}$ are complex fields. 
%%%%%%%%%%%%%%%%%%%%%%%%%%%%%%%%%%%%%%%%%%%%%%%%%%%%%%%%%%%%%%%%%%%%%%%%%%%%%%%%
%%%%%%%%%%%%%%%%%%%%%%%%%%%%%%%%%%%%%%%%%%%%%%%%%%%%%%%%%%%%%%%%%%%%%%%%%%%%%%%%
%% new version
%%%%%%%%%%%%%%%%%%%%%%%%%%%%%%%%%%%%%%%%%%%%%%%%%%%%%%%%%%%%%%%%%%%%%%%%%%%%%%%%
\begin{equation}\label{hyperVariables}
  \big\{q^u\big\}_{u=1}^{4n+4} := \Big(V,a,\xi^A,\xiT_A, z^a,\zb^a\Big)\;,\quad n:=n_H-1\,,
\end{equation}
with $2n+4$ real coordinates $V$,~$a$,~$\xi^A$,~$\xiT_A$,~$A\in\{0\dots n\}$, 
and $n$ complex coordinates $z^a$,~$a\in\{1\dots n\}$.
%%%%%%%%%%%%%%%%%%%%%%%%%%%%%%%%%%%%%%%%%%%%%%%%%%%%%%%%%%%%%%%%%%%%%%%%%%%%%%%%
The scalar $a$ is called the axion since it
has a shift symmetry. In type-IIA 
Calabi-Yau compactifications the scalar $V$ 
is related to the four-dimensional dilaton $\phi_{(4)}$ by $V$~$=$~$\Exp{-2\phi_{(4)}}$,
%parameterises the volume of the Calabi-Yau
%manifold,
and the fields $\xiT_A$,
$\xi^A$ are the massless fluctuations of RR tensor fields.
The complex-structure deformations are conveniently parameterised by scalars $Z^A$, 
which are related
to the fields $z^a$ by
\begin{equation}
  z^A:=(1,z^a)=\big(1,Z^a/Z^0\big)\;.
\end{equation}
In this section we use coordinates $z^A$, while in Section 3 we use
$(Z^A)=Z^0\cdot(1,z^a)$ when lifting the four-dimensional domain-wall solution to 
ten dimensions.%
%\footnote{Essentially, we use $Z^0$ to account for the normalisation of the holomorphic
%  top-form $\Omega$.}
Since we use the c-map for constructing $\Mc_\rmHM$,
the geometry is governed by a
prepotential $F(Z)$. As for the vector multiplet geometry
we take a cubic prepotential:\footnote{Note that the
  factor $i$ in \refeq{F} relative to \refeq{Fc} comes from the fact that we use ``old
  conventions'' of special geometry for $\Mc_\rmHM$ as in Ref.~\cite{FS} and new
  conventions for $\Mc_\rmVM$ as in Ref.~\cite{sugra4d}.}
\begin{equation}\label{F}
  F(Z) := i\;\frac{d_{abc}Z^aZ^bZ^c}{Z^0}\;,
\end{equation}
which corresponds to taking the limit of large complex structures.

If one considers compactifications which are more general
than Calabi-Yau compactifications, the four-dimensional effective action
also contains gauge couplings, a scalar potential and fermionic
mass terms. 
For sake of generality we will first 
consider four-dimensional models with general 
hypermultiplet gaugings, and only later specialise to the particular
gauging occurring in type-IIA compactifications on half-flat six-manifolds.
When starting from an ungauged supergravity action, where all
fields are gauge singlets, a gauging is implemented by 
covariantising the derivatives of those fields which become
charged. In our case these are  only the hypermultiplet scalars $q^u$:
\begin{equation}
  \partial_\mu q^u\quad\longrightarrow\quad
  D_\mu q^u := \partial_\mu q^u + k^ue_IA^I_\mu\;.
\label{gauging}
\end{equation}
The gauge fields $A_\mu^I$ are specific linear combinations of the fields $\Ac_\mu^I$
\cite{sugra4d}. 
It is useful to think about 
the  $e_I$ as deformation parameters which deform 
an ungauged supergravity action into a gauged one.
To have a consistent action of the gauge group on the 
manifold $\Mc_\rmHM$, the gauge group must be a subgroup
of the isometry group of $\Mc_\rmHM$, {\it i.e.}, 
$k^u=k^u(q)$ must be a Killing vector.
Supersymmetry implies in addition 
that the action must now contain a scalar potential and specific
mass terms for the spinors. The general formulae which 
uniquely determine the action in terms of the gauging
(\ref{gauging}) can be found in Ref.~\cite{sugra4d}.
The scalar potential and fermionic mass terms  
contain the so-called Killing-prepotentials $\Pc^1$,~$\Pc^2$,~$\Pc^3$,
which are the quaternionic momentum maps of the isometry $k^u$. Later it
will be convenient to split the $SU(2)$ valued killing prepotential into its norm and a direction:
\begin{equation}\label{killingP}
  \PdS = \sum_{x=1}^3\Pc^x\sigma^x = \sqrt{\PP}\;\nS\;,\quad
  \vec{n}\cdot\vec{n} = 1\;.
\end{equation}
where $\sigma^1$, $\sigma^2$, and  $\sigma^3$ are the Pauli spin matrices.

Supersymmetric vacua $\Phi^0$ %of gauged $\Nc=2$ supergravities 
are found by demanding that 
gravitino, gaugino, and the hyperino variations vanish:
\begin{equation}\label{vars}
  \delta\psi_{\Lambda\mu}\big|_{\Phi^0} = 0\;,\quad
  \delta\lambda^{i\Lambda}\big|_{\Phi^0} = 0\;,\quad
  \delta\zeta_\alpha\big|_{\Phi^0} = 0\;.
\end{equation}
We consider \emph{bosonic}
vacua $\Phi^0$ which implies that the variations of the bosonic fields vanish
identically, since bosons transform into fermions.
Furthermore, we allow only the metric and the scalar fields to depend non-trivially on $y$,
and set $\Ac^I_\mu\big|_{\Phi^0}$~$=$~$0$. We now evaluate the supersymmetry variations
\refeq{vars}, by inserting an Ansatz for the metric and then work out the conditions on
the scalar fields.

We make the following Ansatz for the line element of the domain-wall:
\begin{equation}\label{metric}
 \rmd s^2 = \Exp{2U(y)}\Big[
  \big(\rmd x^0\big)^2-\big(\rmd x^1\big)^2-\big(\rmd x^2\big)^2
  \Big]  -  \Exp{-2pU(y)}(\rmd y)^2.
\end{equation}
Note that we did not fix the transverse coordinate $y$ completely.
The parameter $p$ will be set to a convenient value later, see
Eq.~\refeq{integrability}.

Evaluating the gravitino variation,
we find the following two equations
\begin{eqnarray}
  \label{grav1}
  \partial_y \epsilon_\Lambda &=& -ie^{-pU}\,S_{\Lambda\Delta}\gamma_3\epsilon^\Delta\;,
  \\
  \label{grav2}
  \big(\partial_yU\big)\epsilon_\Lambda &=& 
  -2ie^{-pU}\,S_{\Lambda\Delta}\gamma_3\epsilon^\Delta\;,\quad
\end{eqnarray}
with $S_{\Lambda\Sigma}$~$:=$~$\frac{i}{2}\PdS_\Lambda^{\ \ \Delta}\epsilon_{\Sigma\Delta}
\big(\eL\big)$ \cite{sugra4d}.
%$S_{\Lambda\Sigma} = \frac{i}{2}\PdS_\Lambda^{\ \ \Delta}\epsilon_{\Sigma\Delta}\, \big(\eL\big)
%$
These equations imply that the supersymmetry parameter $\epsilon_\Lambda$ is
proportional to a constant spinor $\epsilon_\Lambda^{(0)}$
\begin{equation}
  \epsilon_\Lambda = \Exp{U/2}\epsilon_\Lambda^{(0)}\;.
\end{equation}
Furthermore, a consistency relation can be derived  by taking the complex
conjugate of Eq.~\refeq{grav2} \cite{BCL}:
\begin{equation}\label{ansatz}
  \epsilon_\Lambda = h(y)\,\nS_\Lambda^{\ \ \Delta}
  \epsilon_{\Sigma\Delta}\,\gamma_3\,\epsilon^\Sigma\;,\qquad
  h\hb=1\;,\quad \vec{n}\cdot\vec{n}=1\;.
\end{equation}
where $\vec{n}$ defined in Eq.~\refeq{killingP}.  This equation generalises the Ansatz
made in Eq.~(3.3) of Ref.~\cite{BCL} to arbitrary hypermultiplet gaugings.  
If \refeq{ansatz} and the Majorana condition 
are the only restrictions on the supersymmetry parameters, then there
are four Killing spinors and we have a 
$\ft12$-BPS background, {\it i.e.}, the domain wall is invariant
under four of the eight real supertransformations of the underlying
supergravity lagrangian.

In the following it will be convenient to define $W$~$:=$~$\sqrt{\PP}\;\big(\eL\big)$.
Inserting Eq.~\refeq{ansatz} into  \refeq{grav2} results in
\begin{equation}\label{hW}
  \hb W = h\Wb = U'\,\Exp{pU}\;,
\end{equation}
and inserting Eq.~\refeq{ansatz} and \refeq{hW} into \refeq{grav1} results in
$\partial_y(h\vec{n})=0$.

The gaugino variation, together with the Ansatz \refeq{ansatz} yields
\begin{equation}
  g_{\jb i}\partial_y z^i = -he^{-pU}\nablab_\jb\Wb\;,
\end{equation}
which precisely is Eq.~(3.15) of Ref.~\cite{BCL}. 
Following their analysis, we find
\begin{equation}\label{YL}
  Y^I := e^U\hb L^I\;,\quad
  \Fc_I := e^U\hb M_I\;,\quad
  i\big(\Yb^I\Fc_I - Y^I\Fcb_I \big) = e^{2U}\;,
\end{equation}
and the following vector multiplet flow equations (VM flow equations):
\begin{equation}\label{stab_eq}
  \frac{\partial}{\partial y}
  \left(\begin{array}{c}
    {Y^I - \Yb^I} \\
    \Fc_I - \Fcb_I
  \end{array}
  \right)
  = -i\sqrt{\PP}\; e^{(1-p)U}
  \left(\begin{array}{c}
    0\\
    e_I
  \end{array}\right)\;.
\end{equation}
We now choose the free parameter $p$ such that
\begin{equation}\label{integrability}
  \sqrt{\PP}\; e^{(1-p)U} = 1\;,
\end{equation}
in order that the flow equations \refeq{stab_eq} can be integrated.
The resulting integrated flow equations
read
\begin{equation}
\label{VM-flow-eq}
\left(\begin{array}{c}
    Y^I - \Yb^I \\
    \Fc_I - \Fcb_I
  \end{array}\right)
  = -i
  \left(\begin{array}{c}
    c^I\\
    H_I
  \end{array}\right)\;,
\end{equation}
where $c^I$ are constants of integration and $H_I$ are harmonic functions
with respect to the flat transverse Laplacian $\partial_y^2$.
We will
refer to both (\ref{stab_eq}) and (\ref{VM-flow-eq})
as VM flow equations in the following, because they characterise
how the vector multiplet scalars evolve as functions of 
the transverse coordinate $y$.
The condition \refeq{integrability} implies the relation
\begin{equation}\label{eY}
  \eY = U'\,\Exp{2U}\in\Rset\;.
\end{equation}
Let us stress that we have derived the flow equations
\refeq{stab_eq}, \refeq{VM-flow-eq} 
without specifying the form of the gauging.

The equations (\ref{VM-flow-eq}) are known as the 
generalised stabilisation equation in the context of
supersymmetric black-hole solutions \cite{Sabra,BLS,CdeWKM}.
The so-called stabilisation equations, which characterise
the attractor behaviour of black hole horizons \cite{FKS}
do not have
a counterpart in our domain wall solution, because the 
underlying lagrangian does not have fully supersymmetric
vacua. In contrast, supersymmetric black hole solutions
interpolate between two fully supersymmetric vacua, 
Minkowski space at infinity and $AdS^2 \times S^2$ at
the horizon \cite{Gibbons,FKS}. Since for our domain walls
moduli stabilisation (fixed point behaviour) does not occur,
we will refer to the equations (\ref{stab_eq}),(\ref{VM-flow-eq})
as the VM flow equations. Note that the upper half of the 
r.h.s.~of the flow equations \refeq{stab_eq} vanishes. Non-vanishing entries
in the upper half would correspond to assigning magnetic charges
to the fields, which we will not consider here (see however
\cite{BCL}).

%These equations also govern black-hole solutions \cite{MohHabil}.
%Since the upper half of the stabilisation equations is zero,
%we have obtained an electric version of the stabilisation equations. 
%This is 
%due to the fact that on the mirror symmetric type-IIB side
%only NS fluxes have been switched on \cite{GLMW}.
%In the next Section we show how these BPS equations are related to the five-dimensional
%stabilization equations \refeq{eq:STAB}.

%which follows from \refeq{Wdef}, \refeq{hW}, \refeq{eY}, and from
%\begin{equation}
%  \hb W = \hb \sqrt{\PP}\,\eL = \sqrt{\PP}\,\Exp{-U}\eY \;.
%\end{equation}

We now turn to the hypermultiplet sector.
For general hypermultiplet gaugings the hyperino variation takes the form
\begin{equation}\label{hyperino}
  \big(\partial_y q^u\big)\Uc_u^{\Lambda\alpha} = 
  \frac{-2iU'}{\PP}\; k^u\Uc_u^{\Sigma\alpha} \PdS_\Sigma^{\ \ \Lambda}\;.
\end{equation}
Here, $\Uc_u^{\Lambda\alpha}\rmd q^u$ are vielbein one-forms on $\Mc_\rmHM$,
and the indices $\Lambda$ and $\alpha$ transform under the 
reduced tangent space group
$SU(2) \times Sp(2n) \subset O(4n)$. 
Explicit expressions for  $\Uc_u^{\Lambda\alpha}\rmd
q^u$ can be found, \eg,~in Ref.~\cite{FS}.

This is all what can be done for general hypermultiplet gaugings.  While the
gravitino and gaugino variations could be evaluated independently of the form 
of the gauging, we need the explicit expressions
for the killing vector $k$ and its associated killing prepotential $\Pc^x$
in order to further exploit \refeq{hyperino}.

Therefore we now specialise our discussion to the gauging 
\begin{equation}\label{killingVs}
  k=\xi^0\partial_a+\partial_{\xiT_0}\;,
\end{equation}
which occurs in the compactification of 
type-IIA string theory on half-flat manifolds \cite{GLMW}. 
Note that this `generalised axion gauging' is different from 
the axion gauging $k=\der_a$ considered in \cite{BCL}.\footnote{
  See \cite{CMDiss} for a domain-wall solution with $k=\der_a$ and an arbitrary number of
  spectator hypermultiplets.
}

The
corresponding killing prepotentials are (in agreement with \cite{D'A})
\begin{equation}\label{killingP2}
  \Pc^1 = \frac{2}{\sqrt{V(zN\zb)}}\;,\quad
  \Pc^2 = 0\;,\quad
  \Pc^3 = -\frac{\xi^0}{V}\;,
\end{equation}
with $(zN\zb)$~$:=$~$\frac{1}{4}z^A\zb^B\partial_A\partial_BF\big(z^C\big)+c.c.$
In solving the hyperino equation, 
we keep all hypermultiplet scalar fields fixed, except $V$ and $z^a$.
Moreover, one finds that $\xi^0$~$\neq$~$0$ 
implies that $U$ and all scalar fields are constant.
Therefore we set $\xi^0$~$=$~$0$.
Note that the supergravity potential
$\Vc$~$\simeq$~$\tfrac{(\xi^0)^2}{V^2}+\tfrac{4}{V(zN\zb)}$ is minimised by $\xi^0=0$.
We find 
\begin{equation}\label{sol2_V}
  V(y) = V_0\Exp{2U(y)}\;,
\end{equation}
and for the scalars $z^a(y)$ we obtain
\begin{equation}\label{sol2_Z}
   z^a = a^a + ib^a\Exp{2U}\;,\quad a^a,\; b^a\in\Rset\;.
\end{equation}
Note that we need to use that the prepotential $F(Z)$
is chosen to be \emph{cubic} in order that the 
hypermultiplet flow equations (HM flow equations) (\ref{hyperino}) decouple 
for the scalars $z^a$, so that we 
have the simple solution displayed above. 
For a cubic prepotential $F(Z)$ one can find 
explicit expressions for the relevant entries in the matrix $N^{-1}$.
We do not know analogous formulae for general prepotentials, and therefore
it is not clear whether the equations decouple for generic prepotentials $F(Z)$.

We now determine the parameter $p$ in the domain-wall line element by solving 
equation \refeq{integrability}.
For the solution \refeq{sol2_V}, \refeq{sol2_Z} we compute
\begin{equation}
  \PP = \frac{4}{V(zN\zb)} = \frac{4}{-2\,V_0\,d_{abc}b^ab^bb^c}\;\Exp{-8U}\;,
\end{equation}
where we have used $\xi^0$~$=$~$0$, and
$(zN\zb)$~$=$~$-2\,d_{abc}b^ab^bb^c\,\Exp{6U}$.
The condition \refeq{integrability},
\begin{equation}
  1= \PP\;\Exp{2(1-p)\,U} = -2\;\frac{\Exp{(-6-2p)\,U}}{V_0\,d_{abc}b^ab^bb^c}\;,
\end{equation}
determines
\begin{equation}
  p = -3\;,
\end{equation}
and gives a condition on the integration constants:
\begin{equation}\label{Vdbbb}
  V_0\,d_{abc}b^ab^bb^c = -2\;,
\end{equation}
which implies
\begin{equation}\label{ZNZb_id}
  (zN\zb)  = \tfrac{4}{V_0}\,\Exp{6U}\;.
\end{equation}

Let us summarise the domain-wall solution corresponding to the generalised 
axion gauging \refeq{killingVs}:
\numparts
\begin{eqnarray}
  \label{BPSDW_1st}
  \rmd s^2 = \Exp{2U}\!\Big[
  \big(\rmd x^0\big)^2-\big(\rmd x^1\big)^2-\big(\rmd x^2\big)^2
  \Big]  -  \Exp{6U}(\rmd y)^2\;,
\\
  \big\{q^u(y)\big\}_{u=1}^{4n+4} = 
  \Big(V(y),\,a,\,\xi^A,\,\xiT_A,\,z^a(y),\,\zb^a(y)\Big)\,,\quad 
  \xi^0=0\;,
%%% HM
\end{eqnarray}
\begin{eqnarray}
  V   &= V_0\,\Exp{2U}\;,\quad
  &V_0\in\Rset\;,
  \\
  z^a &= a^a+ib^a\,\Exp{2U}\;,\quad
  &a^a{},\,b^a\in\Rset\;,
%%% VM
\end{eqnarray}
\begin{eqnarray}
  \Exp{2U} = i\big(\Yb^I\Fc_I - Y^I\Fcb_I \big)\;,\quad
  V_0\,d_{abc}b^ab^bb^c = -2\;,
\end{eqnarray}
\begin{eqnarray}
  \label{BPSDW_last}
  i\partial_y\!\left(\begin{array}{c}
    Y^I - \Yb^I \\
    \Fc_I - \Fcb_I
  \end{array}\right)
  =
  \left(\begin{array}{c}
    0\\
    e_I
  \end{array}\right)\;.
\end{eqnarray}
\endnumparts

%%%%%%%%%%%%%%%%%%%%%%%%%%%%%%%%%%%%%%%%%%%%%%%%%%%%%%%%%%%%%%%%%%%%%%%%%%%%%%%%
\section{Hitchin's Flow Equations}
%%%%%%%%%%%%%%%%%%%%%%%%%%%%%%%%%%%%%%%%%%%%%%%%%%%%%%%%%%%%%%%%%%%%%%%%%%%%%%%%

In the last section we have shown that the four-dimensional gauged
supergravity action, which is obtained by dimensional reduction of
type-IIA superstring theory has the BPS domain wall solution
\refeq{BPSDW_1st}--\refeq{BPSDW_last}. Since the
scalar potential does not allow maximally supersymmetric solutions,
this domain wall is interpreted as the ground state, and 
it should lift to a supersymmetric solution of the underlying
ten-dimensional theory. This is the subject of the present section,
and the key observation needed to relate four-dimensional to 
ten-dimensional physics is that the flow equations which 
determine the dependence of the scalar fields on the transverse
coordinate $y$ are equivalent to Hitchin's flow equations.

%Having constructed a domain-wall solution to the gauging \refeq{killingVs} in the
%last Section, we now show that the four-dimensional domain-wall solution
%lifts to a ten-dimensional domain-wall solution of
%type-IIA supergravity. In this process the four-dimensional BPS equations become Hitchin's
%flow equations.

Let us recall the relevant points of the dimensional reduction 
of type-IIA string theory on half-flat manifolds \cite{GLMW}. For 
Calabi-Yau manifolds one expands the ten-dimensional fields in
terms of a basis of harmonic forms,
\begin{equation}\label{hf_basis}
  \{\alpha_A,\beta^A\}\in\Omega^3(X_6)\;,\qquad
  \omega_i\in\Omega^2(X_6)\;,\quad
  \nu^i\in\Omega^4(X_6)\;,\quad
\end{equation}
where
\begin{equation}\label{hf_basis_i}
  \int_{X_6}\alpha_A\wedge\beta^B = \delta_A^B\;,\quad
  \int_{X_6}\omega_i\wedge\nu^j =   \delta_i^j\;,\quad
  \int_{X_6}\omega_i\wedge\omega_j\wedge\omega_k = c_{ijk}\;.
\end{equation}
The proposal of \cite{GLMW} is that the compactification on a 
half-flat manifold can be treated as a deformation with 
parameters $e_i$, where 
some of the forms cease to be closed (and hence, are not
harmonic any more), 
\begin{equation}\label{hf_basis_d}
  \rmd\alpha_0 = e_i\nu^i\;,\quad
  \rmd\alpha_a = 0 = \rmd\beta^A\;,\quad
  \rmd\omega_i = e_i\beta^0\;,\quad
  \rmd\nu^i=0\;,
\end{equation}
while the relations \refeq{hf_basis_i}
are preserved. Assuming this, the dimensional reduction 
of type-IIA string theory on a half-flat manifold yields 
a four-dimensional supergravity action with the
generalised axion gauging \refeq{killingVs}. 
The deformation parameters $e_i$ in \refeq{hf_basis_d}
are identical with the parameters specifying the
gauging in \refeq{gauging}.
The parameter $e_0$ appearing in
\refeq{gauging} corresponds to switching on an additional flux 
and is set to zero in the following.
It has been shown in \cite{GLMW} that the
relations \refeq{hf_basis_d} 
are fixed by requiring that  half-flat type-IIA compactifications are 
the mirror symmetry partners of type-IIB compactifications on 
(conformally rescaled) Calabi-Yau three-folds with NS-NS flux.

In order to elaborate on the results of \cite{GLMW} and to 
provide a further check of their proposal, we make use
of Hitchin's flow equations. These equations
determine how a family of half-flat
six manifolds $X_6(\yt)$ has to vary along an interval $I$, 
with a coordinate $\yt$ which is
   related to $y$ in \refeq{metric} by the coordinate transformation
  \refeq{m_dy},
 such that one obtains 
a seven-dimensional manifold $Y_7$ with holonomy contained in $G_2$.
Since such manifolds are Ricci-flat, our strategy will be 
to show that the flow of scalar fields along our domain wall
solution is precisely such that a $G_2$ holonomy 
manifold is obtained by combining
the transverse direction of the domain wall solution with the
internal six-manifold.

Recall that the existence of a $G_2$ structure 
on a seven-manifold $Y_7$ requires the existence
of a $G_2$ invariant three-form $\varphi$, while
an $SU(3)$ structure on a six-manifold $X_6$
is equivalent
to the existence of a non-degenerate real two-form $J$ and
of complex three-form $\Omega$, which satisfy \cite{Joyce,ChiossiSalomon}:
\begin{equation}\label{su3eq}
  J\wedge J\wedge J=\tfrac{3i}{4}\,\Omega\wedge\OmegaB\;,\quad
  J\wedge\Omega = 0\;.
\end{equation}
Note that a manifold with $SU(3)$ structure is almost complex,
so that it makes sense to talk about complex differential 
forms.
The complex structure need not be integrable. Therefore 
local complex coordinates need not exist. The real and imaginary
part of $\Omega$ are denoted $\Omega_\pm$. On Calabi-Yau manifolds,
the $SU(3)$ structure is integrable, $\rmd J=\rmd\Omega=0$. In this case $J$ is the K\"ahler
form and $\Omega$ is the holomorphic top form. 

If $Y_7$ is obtained by fibering a family $X_6(\yt)$ of six-manifolds over
an interval $I$, then these data are related by
\cite{ChiossiSalomon}:\footnote{ Note that we have 
rotated $\Omega$ with respect to the convention in
  the mathematical literature, as, \eg,~in Ref.~\cite{ChiossiSalomon}:
  $\big(\psi_+,\psi_-\big)$~$\rightarrow$~$\big(\Omega_-,-\Omega_+\big)$.  }
\numparts
\begin{eqnarray}
  \varphi &= J\wedge\rmd\yt + \Omega_-\;,\\
  \star\varphi &= -\Omega_+\wedge\rmd\yt + \tfrac{1}{2}\,J\wedge J\;.
\end{eqnarray}
\endnumparts
Accordingly, we 
split the exterior derivatives on $Y_7$ into a part inside $X_6$ and the
part along the interval
$I$, $\rmd$~$=$~$\rmdH+\rmd\yt\,\partial_\yt$.
Further details on the relation between $G_2$ structures on seven-manifolds and $SU(3)$
structures on six-manifolds can be found, \eg,~in \cite{KMT,BJ}.

If one demands the stronger condition that $Y_7$ 
has $G_2$ holonomy rather than just $G_2$ structure, 
then $\rmd\varphi=\rmd\star\varphi=0$.
Imposing in addition that $X_6$ is half-flat,
\begin{equation}\label{half-flat-Eq}
  \rmdH\Omega_-=J\wedge\rmdH J=0\;,
\end{equation}
we obtain Hitchin's flow equations:
\numparts
\begin{eqnarray}
  \label{HFE1} 
  \rmdH \Omega_+ &=& \tfrac{1}{2}\,\partial_\yt\big(J\wedge J\big)\;,
  \\
  \label{HFE2} 
  \rmdH J &=& \partial_\yt\Omega_-\;.
\end{eqnarray}
\endnumparts
Hitchin \cite{Hitchin1,Hitchin2} proved that these flow equations preserve the 
$SU(3)$-structure of the fibres \refeq{su3eq}.

We will now show that the flow equations of our domain wall
solution are equivalent to Hitchin's flow equation. 
%Here one
%needs to take into account that the conventions used in the
%supergravity literature and in the mathematical literature
%are slightly different. 
The two- and three-forms 
$(J$,\;$\Omega)$ have an 
expansion in terms of the basis \refeq{hf_basis} as
\begin{eqnarray}
  \label{Jsugra}
  J &=& v^i\omega_i\;,\quad 
  v^i\in\Rset\;,
  \\
  \label{Omega3}
  \Omega&=&\Omega(Z) = \Big(Z^A\alpha_A + \tfrac{i}{2}\,F_A(Z)\beta^A\Big) 
  = Z^0\,\Omega\big(z\big)\;,
\end{eqnarray}
where the moduli $v^i$ are proportional to the supergravity quantities $Y^i$.
Note that $\Omega$ and $\Omega(z)$ differ by the factor $Z^0$, which we fix 
by normalising $\Omega$ as in Eq.~\refeq{su3eq}.
It turns out that the factor $Z^0$ in \refeq{Omega3} is important in matching
the BPS equations to Hitchin's
flow equations. For instance, by computing
\begin{equation}\label{OOb}
\fl  i\int_{X_6}\Omega(z)\wedge\OmegaB(\zb) 
  = \tfrac{1}{2}\,\Big(z^A\Fb_A(\zb) + \zb^AF_A(z)\Big) =
  2\big(zN\zb\big)= \Exp{-K} \simeq \Exp{6U},
\end{equation}
and comparing the result to 
$\int_{X_6}J^3\simeq\Exp{2U}$ one concludes that $Z^0$ has to be a nontrivial function of
$y$ in order for \refeq{su3eq} to hold.
%The forms $(J$,\;$\Omega)$ used in the mathematical literature
%obviously have a different normalisation, because \refeq{su3eq} 
%implies that ${\int_{X_6}\Omega(Z)\wedge\OmegaB(\Zb)}$~$\simeq$~${\int_{X_6}J\wedge J\wedge J}$. 

%In fact, the mathematical conventions are such that
%\begin{equation}
%  \Vol(X_6) = \frac{i}{8}\int_{X_6}\Omega\wedge\OmegaB 
%  = \frac{1}{6}\int_{X_6}J\wedge J\wedge J\;.
%\end{equation}
%Therefore the periods of $\Omega$ have an, albeit trivial, dependence on one of the
%K\"ahler moduli, namely the total volume. In order to disentangle the complex structure
%moduli and the K\"ahler moduli completely, one needs to rescale the periods
%appropriately. In supergravity conventions this is implemented by using a rescaled version
%of $\Omega$, which is denoted $\Omega^\sugra$.
%We only need to determine this normalisation factor for the background we are
%interested in.

A straightforward but somewhat tedious calculation shows 
that the scalar flow equations 
equations \refeq{BPSDW_1st}--\refeq{BPSDW_last} match with 
Hitchin's flow equations \refeq{HFE1}, \refeq{HFE2} if one
includes the following $y$-dependent rescalings:\footnote{The matching 
of the flow equations fixes these rescalings uniquely up to 
$y$-independent factors. These were fixed by imposing natural 
normalisation conditions. In particular, we used this freedom 
to avoid additional constant factors in \refeq{VolumeScalar}
and \refeq{10dMetric}.}
\numparts
\begin{eqnarray}
  \label{m_Y}
  \big\{Y^I\big\} &= \Big\{ Y^0=\tfrac{3i}{4\sqrt{V_0}}\,,\;\; 
  Y^i= \tfrac{1}{4\sqrt{V_0}}\,v^i(y)\Big\}\;,
  \\
  \label{m_dy}
  \rmd\yt &= \big(V_0\big)^{-1/2}\Exp{2U(y)}\,\rmd y\;,
  \\
  \label{m_O}
  \Omega(Z) &= V_0\,\Exp{-2U(y)}\,\Omega(z)\;,\qquad   Z^0(y) = V_0\,\Exp{-2U(y)}\,.
\end{eqnarray}
\endnumparts

Let us verify this claim.
First, we turn to the VM flow equations
\refeq{stab_eq}. Since $Y^i\in\Rset$ and $Y^0=\const$, the upper half of the
flow equations,
\begin{equation}
  i\partial_y\big(Y^I-\Yb^I\big) =0\;,
\label{upperhalf}
\end{equation}
is solved. For the lower half \refeq{m_Y} implies
\begin{eqnarray}
  \nonumber
  i\partial_y\big(\Fc_0-\Fcb_0\big) &=& 0\;,\qquad{\rm since}\quad \Fc_0=\Fcb_0\;,
  \\
  \label{se5d}
  i\partial_y\big(\Fc_i-\Fcb_i\big) 
  &=& \tfrac{1}{2\sqrt{V_0}}\,c_{ijk}\,\partial_y\big(v^jv^k\big) = e_i\;.
\end{eqnarray}
The equations \refeq{se5d} can be integrated and
then take the  characteristic form of VM flow equations for 
{\em five-dimensional} black holes \cite{5dBHs} or 
domain walls \cite{LOSW,BG}:
\begin{equation}\label{stab_eq_5d}
  c_{ijk}v^j(y)v^k(y) = H_i(y)\;,\quad
  H_i(y) = 2\sqrt{V_0}\,e_i\;y + \mathrm{const}\;.
\end{equation}
with harmonic functions $\partial_y^2H_i(y)=0$.
The fact that our four-dimensional VM flow equations
can be cast into five-dimensional form reflects that one
can lift our four-dimensional domain wall solution to 
a five-dimensional one.
This has two reasons: first, we have taken the prepotential
to be cubic, and therefore the four-dimensional action
can be obtained by dimensional reduction of a five-dimensional
one. Second, only one real scalar in each four-dimensional
VM flows along the domain wall, namely the 
real part of $Y^I$, which corresponds 
to deformations of the almost-K\"ahler form $J$,
see \refeq{upperhalf}, \refeq{stab_eq_5d}. This field remains
a scalar when lifting the action to five dimensions, while 
the other real scalar (the imaginary part of $Y^I$) 
becomes the fifth component of a gauge field. 
At the level of the 
full ten-dimensional type-IIA string theory this dimensional lift corresponds 
to the M-theory limit. In particular, it is clear that when 
using the basis \refeq{hf_basis} - \refeq{hf_basis_d} for
the dimensional reduction of 
eleven-dimensional supergravity on half-flat six-folds, the
scalars in the vector multiplets correspond to deformations of $J$.

We now observe that 
the first Hitchin flow equation \refeq{HFE1} becomes
\begin{equation}\label{1m}
  V_0\Exp{-2U} e_i\nu^i = 
  \tfrac{\sqrt{V_0}}{2}\,\Exp{-2U}\,c_{ijk}\,\partial_y \big(v^jv^k\big)\nu^k\;,
\end{equation}
when using the basis \refeq{hf_basis} and
\begin{equation}
  \rmdH\Omega(z) = \rmdH\Omega(z)_+ = e_i\nu^i\;,\quad
  J\wedge J = c_{ijk}v^jv^k\nu^i\;.
\end{equation}
In summary the VM flow equation \refeq{stab_eq}, can be rewritten in the
form \refeq{stab_eq_5d} and matches the first Hitchin flow equation \refeq{HFE1}.

We now turn to the second Hitchin flow equation
\refeq{HFE2}. Substituting \refeq{m_dy}, \refeq{m_O}, we obtain
\begin{equation}\label{2m}
  e_iv^i\,\beta^0 = V_0^{3/2}\,\Exp{-4U}\,\Big[ 
  \partial_y\Omega(z)_- - 2\,\big(\partial_yU\big)\,\Omega(z)_- 
  \Big]\,.
\end{equation}
The supergravity three-form fulfils \refeq{Omega3}
\begin{equation}
  \partial_y\Omega(z)_- = 2\big(\partial_yU\big)\,\Omega(z)_- 
  - 2\big(\partial_yU\big)\,\Exp{6U}\,d_{abc}b^ab^bb^c\,\beta^0\;,
\end{equation}
where the last term can be cast into a from compatible with \refeq{2m}
\begin{equation}
  -2\big(\partial_yU\big)\Exp{6U}\,d_{abc}b^ab^bb^c =  V_0^{-3/2}\Exp{4U}\,e_iv^i\;,
\end{equation}
by making use of Eq.~\refeq{eY} and Eq.~\refeq{Vdbbb}. Therefore, the second
Hitchin flow equation \refeq{HFE2} matches the HM flow equations.

Finally, let us show that the 
supergravity solution is compatible with 
the normalisation used in the $SU(3)$ structure equations \refeq{su3eq}.
We have 
already shown
that the scalar flow equations imply Hitchin's flow equations, but
this only guarantees that \refeq{su3eq} is preserved under the 
flow. 
%Since according to \refeq{m_J} and \refeq{m_O} $(J,\Omega)$ 
%and $(J,\Omega)(z)$ only differ by a $y$-dependent factor, 
All we need to check is that $J$ and $\Omega$
are related by $J\wedge J \wedge J = \ft{3i}4\,\Omega\wedge\OmegaB$ 
if we plug in the supergravity solution, taking in mind that $\Omega=\Omega(Z)$ is related
to $\Omega(z)$ by the factor $Z^0(y)$ given in \refeq{m_O}.
We find
\begin{equation}\label{norm1}
  \Vol\big(X_6\big) = \frac{1}{6}\,\int_{X_6}J\wedge J\wedge J = \tfrac16\,c_{ijk}v^iv^jv^k 
  = V_0\,\Exp{2U}\;,
\end{equation}
where we have used $e^{2U}$~$=$~$\tfrac{1}{6V_0} c_{ijk}v^iv^jv^k$. Similarly we obtain
\begin{equation}\label{norm2}
\fl  \Vol\big(X_6\big) = \frac{1}{6}\frac{3i}{4}\,\int_{X_6}\Omega(Z)\wedge\OmegaB(\Zb) 
  = Z^0\Zb^0\,(zN\zb)/4
%%  = \frac{V_0^2\,(zN\zb)}{4}\;\Exp{-4U}
  = V_0\,\Exp{2U}\;,
\end{equation}
using Eq.~\refeq{ZNZb_id} and Eq.~\refeq{OOb}.
As expected, Eq.~\refeq{norm1} and Eq.~\refeq{norm2} match, and
the volume of $X_6$ is
parameterised by the hypermultiplet scalar $V$,
\begin{equation}
\label{VolumeScalar}
  \Vol\big(X_6\big) = V_0\,\Exp{2U} = V(y) = \Exp{-2\phi_{(4)}(y)}\;,
\end{equation}
where $\phi_{(4)}$ is the four-dimensional dilaton.

Finally, let us specify the ten-dimensional line element
explicitly. The ten-dimensional and 
four-dimensional line elements are related by:\footnote{
This relation holds in general for the dimensional reduction of a
ten-dimensional gravity action on a six-manifold $X_6$, if one
parametrises both the ten-dimensional and the 
four-dimensional action such that the Einstein-Hilbert terms take
their canonical form (`Einstein frame'). See, \eg,~\cite{Bodner:1991}.}
\begin{equation}
\rmd s^2_{(4)} = \Vol\big(X_6\big)\; \rmd s^2_{(10)} =
  V_0\,\Exp{2U}\;  \rmd s^2_{(10)}\;.
\end{equation}
The four-dimensional line-element is
\begin{equation}
    \rmd s^2_{(4)}
    = e^{2U}\big(\rmd s^2\big)_{1,2} - e^{6U}(\rmd y)^2
    = e^{2U}\,\Big[\big(\rmd s^2\big)_{1,2} - V_0\,\big(\rmd\yt\big)^2\Big]\;.
\end{equation}
Hence, the ten-dimensional line element is given by
\begin{equation}
\label{10dMetric}
  \rmd s^2_{(10)} =  \Vol\big(X_6\big)^{-1}\;\rmd s^2_{(4)} - e^a\otimes e^a 
  = \tfrac{1}{V_0}\,\big(\rmd s^2\big)_{1,2} -
  \big(\rmd\yt\big)^2 - e^a\otimes e^a\;,
\end{equation}
where $e^a$ is the $y$-dependent vielbein of the internal six-dimensional
half-flat manifold.
By rewriting the metric in terms of the rescaled coordinate $\yt$  
we see that the ten-dimensional metric factorises into a flat
three-dimensional Minkowski space and a seven-dimensional part
\begin{equation}
X_{10} =   \Rset^{1,2}\times Y_7\;.
\end{equation}
The seven-manifold $Y_7$ with line-element $\big(\rmd\yt\big)^2 + e^a\otimes e^a$
is a fibration over an interval. Since the $\yt$-dependence
is given by Hitchin's flow equations \refeq{HFE1}, \refeq{HFE2},
and since the six-dimensional part is half-flat,
the  holonomy group is contained in $G_2$, and $Y_7$ is a
$G_2$-holonomy manifold with boundaries. Thus $Y_7$ and
$X_{10}$ are Ricci-flat. 
Since all non-constant
scalar fields of our four-dimensional domain wall come
from internal components of the ten-dimensional metric, 
the ten-dimensional type-IIA supergravity equations of
motion are satisfied.

It is easy to see why the scalar $V$~$=$~$\Exp{-2\phi_{(4)}}$, where $\phi_{(4)}$ is the
four-dimensional dilaton is equal to the volume of $X_6$. Recall that the ten-dimensional
and the four-dimensional dilaton are related by 
\begin{equation}
  \Exp{-2\phi_{(4)}} = \Exp{-2\phi_{(10)}}\,\Vol(X_6)\;.
\end{equation}
In our solution all ten-dimensional mater fields including the ten-dimensional dilaton
are trivial and therefore $\phi_{(10)}$~$=$~const. Taking $\phi_{(10)}$~$=$~$0$ for
convenience\footnote{
  Any other value of $\phi_{(10)}$ just leads to an additional irrelevant constant,
  which can be absorbed by rescaling the ten-dimensional gravitational constant.
  For $\phi_{(10)}$~$=$~$0$ the ten-dimensional string frame and Einstein frame coincide.
  Note that the four-dimensional string and Einstein frame differ, because
  $\phi_{(4)}$~$\neq$~const. All our four-dimensional metrics refer to the Einstein frame.
%%  Taking a different vacuum expectation value for $\phi_{(10)}$ can be absorbed by rescaling
%%  the ten-dimensional coupling. Alternatively, we could keep the vacuum expectation value
%%  of $\phi_{(10)}$ explicitely, but this would only lead to an irrelevant constant.
} we see 
that the four-dimensional dilaton is related to the volume of $X_6$ as stated above.

%%------------------------------------------------------------------------------
%%%%%%%%%%%%%%%%%%%%%%%%%%%%%%%%%%%%%%%%%%%%%%%%%%%%%%%%%%%%%%%%%%%%%%%%%%%%%%%%
\section{Conclusions}
%%%%%%%%%%%%%%%%%%%%%%%%%%%%%%%%%%%%%%%%%%%%%%%%%%%%%%%%%%%%%%%%%%%%%%%%%%%%%%%%

In this paper we have analysed domain-wall solutions of gauged four-dimensional $\Nc=2$
supergravity with arbitrary hypermultiplet gaugings. We have found that the 
VM flow equations \refeq{VM-flow-eq} are universal, while the HM flow
equations depend on the precise form of the gauging.

Furthermore, we have constructed a new type of domain wall solution
\refeq{BPSDW_1st}--\refeq{BPSDW_last},
which
generalises the domain walls of \cite{BCL} 
from the universal hypermultiplet to an arbitrary number of hypermultiplets, which all
vary non-trivially. In addition, we considered a different gauging, the 
`generalised axion shift' $k=\xi^0\partial_a+\partial_{\xiT_0}$, while in Ref.~\cite{BCL}
the axion shift symmetry, $k=\partial_a$, was used.
A domain wall
solution with the axion shift and spectator hypermultiplets can be found in \cite{CMDiss}.

Our domain walls are the vacua of a four-dimensional supergravity theory which was obtained by
dimensional reduction of type-IIA supergravity on half-flat manifolds \cite{GLMW}.
In Section 3, we have shown that, as anticipated by \cite{GLMW}, the
four-dimensional domain-walls lift to ten-dimensional backgrounds of
type-IIA string theory. The ten-dimensional geometry is of
the form $\Rset^{1,2}$~$\times$~$Y_7$ where $Y_7$ is a $G_2$-holonomy manifold.
The four-dimensional flow equations become  Hitchin's flow equations which determine
how the family of half-flat manifolds $X_6(\yt)$ makes up the $G_2$ holonomy manifold
$Y_7$.

Since the action of \cite{GLMW} is manifestly mirror symmetric, it is clear that  our
domain walls are also solutions of type-IIB string theory compactified on a Calabi-Yau
three-fold with NS flux. Therefore, we expect that it can be lifted to a consistent
background of the full ten-dimensional type-IIB string theory.
It would be interesting to perform an analysis of such a type-IIB background as explicitly
as for the type-IIA background in this paper.

In this context one could ask if there are further type-II string geometries, 
where in the four-dimensional effective action other isometries than the axion and the
generalised axion shift symmetries are gauged. 
%there are consistent type-II string theory
%backgrounds of the form $\Rset^{1,3}$~$\times$~$X_6$ with background flux.
A first step is to construct four-dimensional domain-wall solutions
corresponding to other gaugings than we have used here. All
what has to be done is to solve the hyperino variation, \refeq{hyperino},
since the VM flow equations take the same form for all
hypermultiplet gaugings.  These new domain-walls will then be lifted to ten
dimensions. We expect that in this context the generalised
Hitchin's flow equations derived in \cite{D'A2} play a role, and that one
obtains a manifold with $G_2$ structure. These generalised
flow equations can be derived similarly to Hitchin's
flow equations with the difference that now there are non-vanishing torsion classes
on the right hand side:
\numparts
\begin{eqnarray}
  \label{1st}
  \rmd\star\varphi &=\dots
  \quad\longleftrightarrow\quad
  {\rm VM~flow~equations}
  \\
  \label{2nd}
  \rmd\varphi &= \dots
  \quad\longleftrightarrow\quad
  {\rm HM~flow~equations}
\end{eqnarray}
\endnumparts
Perhaps one of the most intriguing points is that the vector
multiplet flow equations are universal. This may be used to derive constraints
on the $G_2$ torsion classes, since we know that the VM
flow equations come from \refeq{1st}.

Phenomenologically, it would be interesting to
lift our setup by one dimension to a M-theory background of the form
$\Rset^{1,3}$~$\times$~$Y_7$. If the $G_2$-holonomy manifold $Y_7$ has no boundaries one would
recover four-dimensional $\Nc$~$=$~$1$ compactification of M-theory \cite{Acharya}. In the
other case,  
if $Y_7$ has boundaries, models of Ho\v{r}ava-Witten-type \cite{HW, LOSW} would result. 
Alternatively, one can ask if it is possible to have string backgrounds of the form
$\Rset^{1,3}$~$\times$~$X_6$ where  the curvature of $X_6$ is balanced by background flux.

Another interesting aspect is how $SU(3)$ 
structures are related to gaugings of four-dimensional supergravity actions.
For instance, it would be interesting to obtain a map
between $SU(3)$ torsion classes and hypermultiplet gaugings.

\ack

We would like to thank K.~Behrndt, G.~L.~Cardoso, T.~Grimm, J.~Louis, D.~L\"ust, and
A.~Micu for useful discussions.  This work is supported by the `Schwerpunktprogramm
Stringtheorie' of the DFG.

\end{document}